\documentclass[%
 reprint,
 bibnotes,
 amsmath,amssymb,
 aps,
 prl,
]{revtex4-2}

\usepackage{graphicx}
\usepackage{dcolumn}
\usepackage{bm}
\usepackage{hyperref}

\begin{document}

\preprint{APS/123-QED}

\title{\textbf{Multiple self-organized phases and spatial solitons in cold atoms mediated by optical feedback}}

\author{Giuseppe Baio}
\email{giuseppe.baio@strath.ac.uk}
\author{Gordon R. M. Robb}
\author{Alison M. Yao}
\author{Gian-Luca Oppo}
\author{Thorsten Ackemann \vspace{.1cm}}

\affiliation{
 SUPA and Department of Physics, University of Strathclyde, Glasgow G4 0NG, Scotland, United Kingdom
}%

\date{\today}

\begin{abstract}
We study the transverse self-structuring of a cloud of cold atoms with effective atomic interactions mediated by a coherent driving beam retro-reflected by means of a single mirror.
The resulting self-structuring due to optomechanical forces is much richer than that of an effective-Kerr medium, displaying hexagonal, stripe and honeycomb phases depending on the interaction strength parametrized by the linear susceptibility. Phase domains are described by real Ginzburg-Landau amplitude equations. In the stripe phase the system recovers inversion symmetry. Moreover, the subcritical character of the honeycomb phase allows for light-density feedback solitons functioning as self-sustained dark atomic traps with motion controlled by phase gradients in the driving beam.
\end{abstract}

\maketitle


\newcommand{\rp}{\mathbf{r}}
\newcommand{\fdip}{\mathbf{f}_{\textrm{dip}}}
\newcommand{\nabp}{\nabla_{\perp}}

Spontaneous self-organization phenomena are ubiquitous in out-of-equilibrium classical and quantum dynamics \cite{Cross1993, polkovnikov2011colloquium}. In recent years, cold and ultracold gases have provided useful platforms for probing light-atom self-structuring by means of density modes or internal states, resulting in crystalline (density) or magnetic order respectively \cite{Labeyrie2014a, Ostermann2016, labeyrie2018magnetic, kresicCP, landini}. In the first case, the emerging dynamical potential for the atoms induces a density grating which, in turn, scatters photons into the side-band modes creating the potential and leading to optomechanical self-structuring \cite{Ritsch2013}. Several experimental realizations of such phenomena in cold atoms setups have offered groundbreaking insight into different quantum many-body physics aspects such as quantum phase transitions \cite{baumann2010dicke}, supersolidity \cite{gopalakrishnan09,gopalakrishnan10, mottl12, leonard17}, topological defects \cite{Labeyrie2016}, and structural phase transitions \cite{li2020measuring}.

A key aspect, analogous to soft-matter realizations \cite{reece2007experimental}, is that the collective bunching of the scatterers gives rise to a self-focusing Kerr-like optomechanical nonlinearity \cite{gupta2007cavity}. Hence, Ashkin \textit{et al.} coined the term `artificial Kerr medium' \cite{ashkin82}. Transverse optical pattern formation in effective-Kerr media (and beyond) has been the subject of wide theoretical and experimental efforts since the 1990's, in both cavity and single-feedback-mirror (SFM) configurations \cite{Firth1990, d1991spontaneous, DAlessandro1992, scroggie1994pattern}. Among the major advantages of cold atoms is the possibility to significantly reduce threshold intensities when the atoms are laser cooled to hundreds of $\mu \textrm{K}$ \cite{Labeyrie2014a, Saffbook, Greenberg2011}.

In this Letter, we show that, despite some similarities between Kerr media and mobile dielectric scatterers, the latter is source of a much richer structural transition behaviour characterized by three light-atom crystalline phases, i.e., hexagonal, stripe and honeycomb. We explore phase stability for an SFM setup in terms of a weakly nonlinear analysis, leading to the amplitude equations (AEs) and relative free energy functional in the universal Ginzburg-Landau form \cite{CGLE}. This provides an accurate description of the selection mechanism and spatial soliton formation in a cloud of atoms undergoing optomechanical self-structuring. Our results can be applied to other configurations of interest, e.g, in free-space or longitudinally pumped cavities \cite{Schmittberger2016, Tesio2012}, and can shed new light on the ongoing discussion of potential phases in the rapidly developing field of dipolar supersolids \cite{tanzi2019observation, bottcher2019transient, chomaz2019long}. Indeed, although current experimental realizations are limited to quasi-1D cases, 2D structural transitions are predicted to occur in dipolar condensates \cite{Zhang2019}. Based on a close correspondence between the condensate energy functional and the Lyapunov functional discussed below, we conjecture that our analysis will motivate further studies in the potential of a new stripe supersolid phase in between the hexagonal and honeycomb phases already predicted in  \cite{Zhang2019}. 

\begin{figure}
\hspace*{.5cm}\includegraphics[scale=1.56]{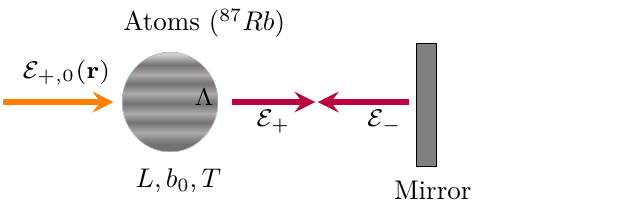}
\vspace{-.5cm}
\caption{\footnotesize Optomechanical SFM scheme. A far detuned input beam of amplitude $\mathcal{E}_{+,0}$ and wavenumber $k_{0}$ illuminates a cloud of Rubidium atoms of thickness $L$, optical density $b_0$ and temperature $T$. The reflected $\mathcal{E}_{-}$ provides feedback by means of the dipole potential, leading to self-structuring with critical wavelength $\Lambda$ \citep{Labeyrie2014a}.}
\label{sfm}
\end{figure}

\begin{figure}
    \centering
    \hspace*{-0cm}\includegraphics[scale = .148]{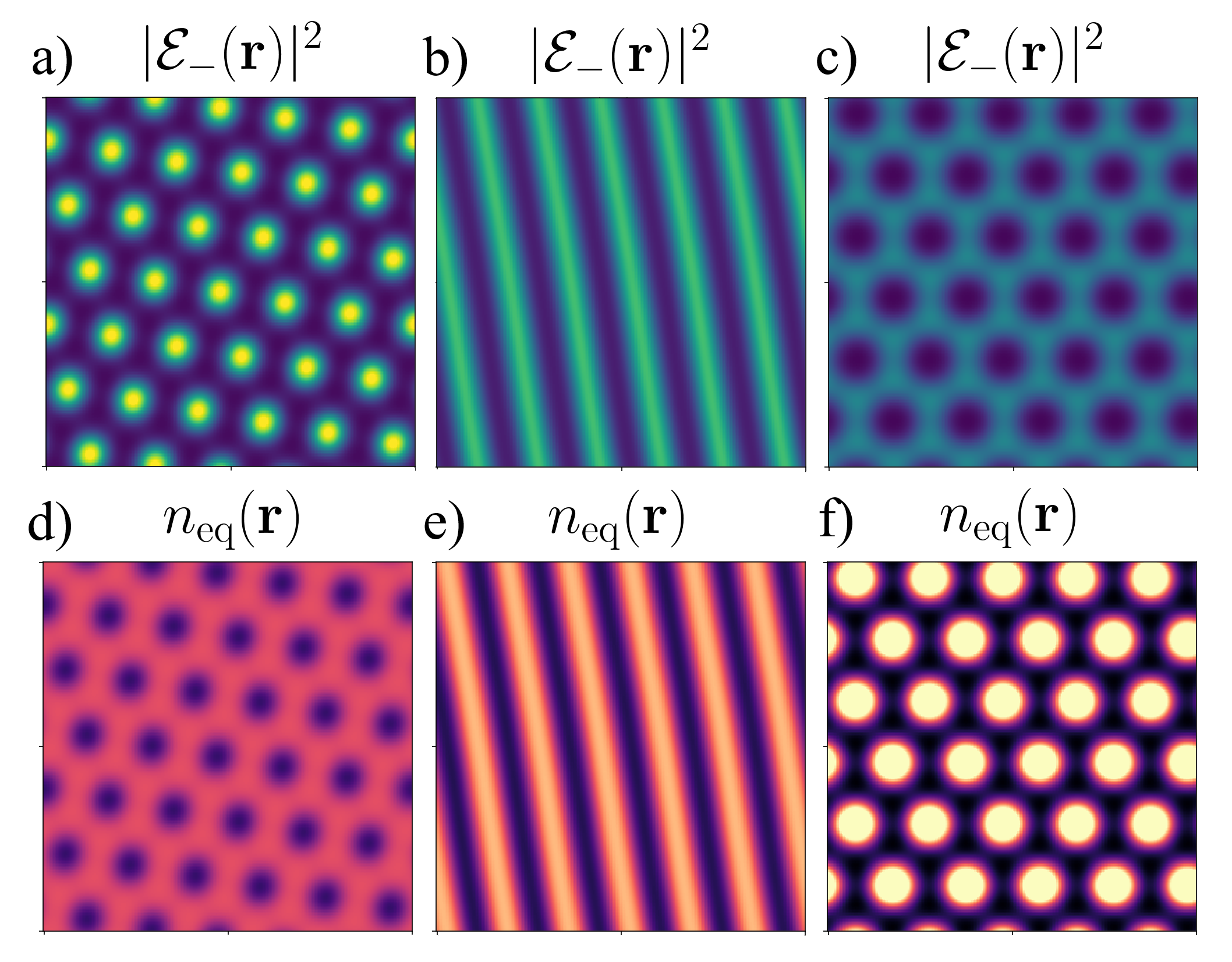}
    \caption{Optomechanical self-structured phases obtained at fixed $b_0 = 110$ and $T=300\,\, \mu\textrm{K}$.  (a),(d) $\mathbf{H}^-$ phase at $\Delta = 25$. (b),(e) $\mathbf{S}$ phase at $\Delta = 55$. (b),(e) $\mathbf{H}^+$ phase at $\Delta = 90$.  }
    \label{collage}
\end{figure}

\begin{figure}
    \centering
    \hspace*{-.5cm}\includegraphics[scale = .125]{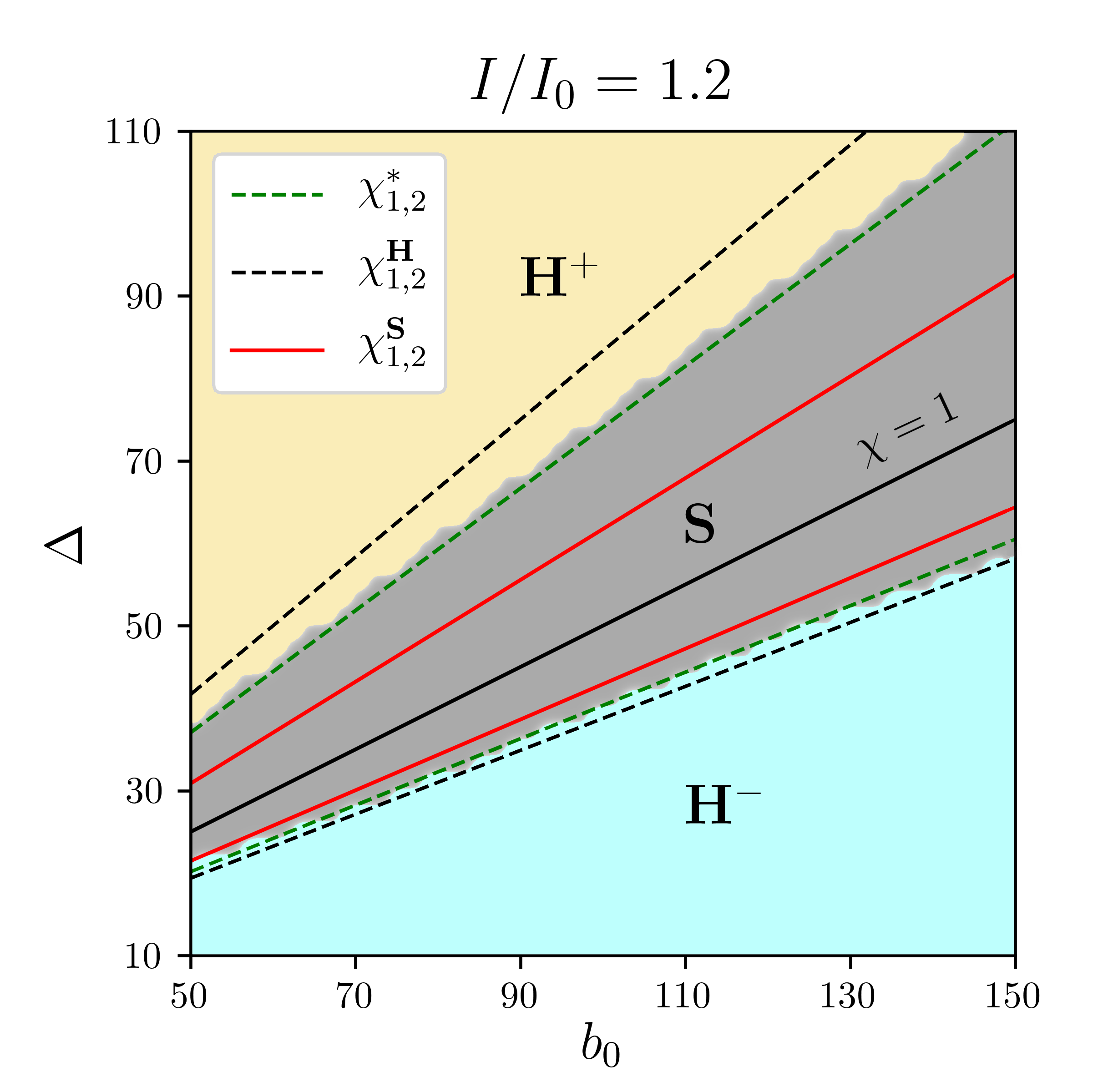}
    \caption{Numerically observed stability domains of the $\mathbf{S}$, $\mathbf{H}^\pm$ phases at fixed $I/I_0$. The observed boundaries match the values of the susceptibility $\chi$ from Eqs. (\ref{lyap})-(\ref{crit}). The $\mathbf{S}$ phase (grey) is absolutely stable on a domain sandwiched between the lines corresponding to the $\chi^{\mathbf{S}}_{1,2}$ points (red), around $\chi=1$ (black). Moreover, the $\mathbf{S}$ phase coexists with $\mathbf{H}^\pm$ and minimizes the free energy in the region between the $\chi^{\mathbf{H}}_{1,2}$ (dashed-black) and $\chi^*_{1,2}$ points (dashed-green). $\mathbf{H}^\pm$ phases are stable within the yellow and cyan domains and absolutely stable outside $\chi^{\mathbf{H}}_{1,2}$.}
    \label{pd20}
\end{figure}

We consider a thermal cloud of two-level atoms at temperature $T$, where atomic motion is overdamped by means of optical molasses \cite{Saffbook}. In this regime, the transverse dynamics is described by density modulations only, i.e.,  $n(\rp,t) = 1 + \delta n(\rp,t)$, where the atom density $n(\rp,t)$ obeys a Smoluchowski drift-diffusion equation \citep{Tesio2012, Ritsch2013}:
\begin{equation}
\partial_t n(\rp,t)= -\beta D\nabp\cdot\left[n(\rp,t)\,\fdip(\rp,t)\right]+D\nabp^{2}n(\rp,t),
\label{smolu}
\end{equation}
where $D$ is the cloud diffusivity, and $\beta = 1/k_{B}T$, with $k_{B}$ being the Boltzmann constant. The dipole force reads: \begin{equation}
\fdip(\rp,t)  =  -\frac{\hbar\Gamma\Delta}{4}\nabp s(\rp,t),
\label{dipolf}
\end{equation}
where $s(\rp,t) $ is the total light intensity (saturation parameter), and $\Delta$ corresponds to the light-atom detuning in units of half the linewidth  $\Gamma$. The SFM setup, represented in Fig.~\ref{sfm}, is a paradigmatic scheme for Talbot-based optical pattern formation \cite{Firth1990, DAlessandro1992, Scroggie1996}. For a diffractively thin cloud, the field equations are:
\begin{equation}
\partial_z \mathcal{E}_{\pm}(\rp,t) = \pm i\frac{\chi}{L} n(\rp,t)\,\mathcal{E}_{\pm}(\rp,t),
\label{field}
\end{equation}
where the $+$ sign relates to $\mathcal{E}_{+}$ and vice versa.  As typical for the dispersive regime, we assume large detuning and low saturation, so that scattering forces are neglected and the susceptibility of the cloud is real and reads $
\chi = b_0\,\Delta/2(1+\Delta^2)
\label{chi}
$ (See Fig.~\ref{sfm}) \cite{Labeyrie2014a}.
The feedback loop is closed by considering propagation to the mirror (at distance $d$ from the cloud) and back \cite{Ackemann2001}. Let us introduce a constant $\sigma = \hbar\Gamma\Delta/4k_{B}T$, representing competition between the dipole potential and spatial diffusion. We first study the linear stability of spatial modulations \cite{Note1}. By parametrizing $\delta n(\mathbf{q},t) = a\exp(i\mathbf{q}\cdot\mathbf{r}+\nu t) + \textrm{c.c.}$, one obtains the following  growth rate:
\begin{equation}
\nu(\mathbf{q}) = -D|\mathbf{q}|^2\left[ 1-  \frac{\sigma R|\mathcal{E}_{+,0}|^2 b_0\,\Delta \sin\Theta}{(1+\Delta^2)} \right],
\end{equation}
where $\Theta = d|\mathbf{q}|^2/k_0$ is the total diffractive phase slippage, $R$ is the mirror reflectivity and $\mathbf{q}$ is the transverse wavevector. Imposing  $\nu(\mathbf{q}) = 0$, we arrive at the threshold condition:
\begin{equation}
I = |\mathcal{E}_{+,0}|^2 = \frac{1+\Delta^2}{\sigma R\, b_0\, \Delta \sin\Theta} \geq \frac{1+\Delta^2}{\sigma R\, b_0\, \Delta } = I_0,
\label{threshold}
\end{equation}
where $I_0$ represents  the minimum threshold, i.e.,  at the critical wavenumber $ q^2_c = k_0\pi/2d $ (purely dispersive case). We explore the coexistence of self-structured phases by means of numerical and analytical observations. Unlike the molasses-free case considered in  \cite{Tesio2014a},  the dissipative dynamics of Eq. (\ref{smolu}) admits a (quasi) stationary state given by the Gibbs distribution \cite{Tesio2012}:
\begin{equation}
n_{\textrm{eq}}(\mathbf{r}, t)  =\frac{\exp[-\sigma s(\mathbf{r},t)]}{\int_{\Omega}d^2\mathbf{r}\exp[-\sigma s(\mathbf{r},t)]},
\label{gibbs}
\end{equation}
where $\Omega$ is the integration domain and $s(\rp,t) = |\mathcal{E}_{+}(\rp,t)|^2 + |\mathcal{E}_{-}(\rp,t)|^2 $. The feedback loop is integrated according to the following scheme: first, we propagate the incident field through the cloud, i.e., $ \mathcal{E}_{+}(z=L,\rp,t) = \mathcal{E}_{+,0}\exp{\left\{ i\chi n(\rp,t) \right\}}$. We then propagate in free-space over $2d$ to determine $\mathcal{E}_{-}(z=L,\rp,t)$, and update the atom density according to $n_{\textrm{eq}}(\mathbf{r}, t)$ in Eq. (\ref{gibbs}). By expanding Eq. (\ref{gibbs}) to first order, one shows that, regardless of the sign of $\chi$, the total refractive index of the cloud is of self-focusing Kerr type and, thus, the atom density is expected to choose an hexagonal (honeycomb) geometry above threshold for $\Delta < 0 $ ($\Delta > 0$) \citep{Labeyrie2014a, DAlessandro1992}. However, for the optomechanical interaction we numerically observe the formation of three self-structured phases shown in Fig.~\ref{collage} for different values of $\Delta$ at fixed $b_0$, i.e., hexagons ($\mathbf{H}^+$), stripes ($\mathbf{S}$) and honeycombs ($\mathbf{H}^{-}$), where the labels identify the atom-density states. To characterize transitions between such phases we span the two dimensional space $(\Delta,b_0)$ within the experimentally achievable ranges of $\Delta=[10,110]$ and $b_0=[50,150]$ \cite{Labeyrie2014a}. The stability diagram shown in Fig.~\ref{pd20} is obtained from numerical simulations by seeding with an $\mathbf{S}$ state and iterating the loop long enough to let the stucture stabilize. A simple discriminant between phases is the number of peaks in the resonant circle of the far field.
In Fig.~\ref{pd20}, we report a stability domain of $\mathbf{S}$ states (in grey) for $I/I_0=1.2$, sandwiched between two disjoint $\mathbf{H}^\pm$ regions (in yellow/cyan) and separated by lines of constant $\chi$.

A weakly nonlinear analysis based on the AEs represents the canonical approach to describe pattern selection processes above threshold \cite{hoyle2006}. The first step is to formally integrate Eq. (\ref{field}) for the backwards field with a homogenous pump, namely:
\begin{equation}
 \mathcal{E}_{-}(\rp,t) = \sqrt{RI}\hat{\mathcal{L}}e^{i\chi n(\rp,t)},
\end{equation}
where we defined the the differential operator $\hat{\mathcal{L}} = e^{-id \nabp^{2}/k_0}$. Thus, we are left with one equation for the density perturbation $ \delta n(\rp,t)$ only:
\begin{equation}\begin{split}
(-\nabp^2&+\partial_t)  \chi \delta n(\rp,t) = \\
&R\sigma I\chi \nabp\cdot\left[(1+\delta n(\rp,t))\,\nabp |\hat{\mathcal{L}}e^{i\chi \delta n(\rp,t)}|^2\right].
\label{exac}
\end{split}\end{equation}

A similar approach was used to derive a closed equation capturing the features of the long-range interactions mediated by feedback in a SFM scheme  \cite{Zhang2018}. We now expand  $|\hat{\mathcal{L}}e^{i\chi \delta n(\rp,t)}|^2$ up to $O\left[(\chi\delta n)^3\right]$ and introduce slow spatial scales up to third order \cite{manneville1990}. Furthermore, we derive the solvability conditions for our model and substitute a hexagonal ansatz for the resonant terms $n_1$:
\begin{equation}
n_1 = \frac12 \left[\sum^3_{i=1} A_i \exp\left(i\mathbf{q}_i\cdot \rp\right)+ \textrm{c.c.}\right],
\end{equation}
where $\mathbf{q}_1+\mathbf{q}_2+\mathbf{q}_3 = 0$ and $|\mathbf{q}_i|=q^2_c$.  After lengthy algebra \cite{Note1}, we obtain the AEs in the real Ginzburg-Landau form, namely:
\begin{equation}
\partial_t A_i = \mu A_i + \lambda A^*_j A^*_k - \gamma_{1}\sum_{j\neq i}|A_j|^2A_i -  \gamma_{2}|A_i|^2A_i.
\label{GLEs}
\end{equation}
where $i, j, k = 1,2,3$ and $i\neq j \neq k$. In many circumstances, pattern stability close to threshold is universally described in terms of the AEs critical points, depending on the coefficients in Eq. (\ref{GLEs}) \cite{DAlessandro1992, Scroggie1996}. Defining $p = I/I_0$, and for a generic critical shift $\Theta_c$,  the linear growth and three-mode mixing coefficients read:
\begin{align}
&\mu(p) =  2 R I_0 \sigma \,(p- 1)  \chi \sin{\Theta_c}, \label{coef11}\\
&\lambda(p,\chi) =  \frac{ R I_0\sigma p \chi}{2}\,  \left[\sin{\Theta_c}+\chi(\cos\Theta_c-1)\right].
\label{coef12}
\end{align}
Already at this level, a number of interesting considerations arise. Indeed, in sharp contrast with the Kerr model, the coefficient $\lambda$ changes sign around the point $\chi = \cot(\Theta_c/2)$ ($\chi =1 $ with $\Theta_c = \pi/2$), determining a change in the type of hexagons observed ($\mathbf{H}^+$ for $\lambda > 0$ and vice versa) \cite{DAlessandro1992}. Secondly, such a change occurs only for $\chi>0$, i.e., for blue-detuning ($\Delta >0$) while, instead, no phase other than $\mathbf{H}^-$ is expected at threshold for red-detuning. As in the Hamiltonian case, phase selection processes, such as the one in Fig.~\ref{pd20}, are described in terms of Lyapunov or free energy functionals associated with the AEs in Eq. (\ref{GLEs}) \cite{hoyle2006}. To this aim, we compute the self and cross-cubic coefficients as follows:
\begin{align}
&\gamma_{1}(p,\chi) =  \frac{ R I_0\sigma p \chi^2}{4} \left[\chi \sin\Theta_c+2 -\frac{1}{2}\left(\cos3\Theta_c+\cos\Theta_c\right) \right] \label{coef21}\\
&\gamma_{2}(p,\chi) =  \frac{ R I_0\sigma p \chi^2}{8}\,\left[\chi\left(\sin\Theta_c-\sin 3\Theta_c+\frac{2}{3}\right)\right.+ \nonumber\\
&\left.2(1-\cos4\Theta_c)\right],
\label{coef22}
\end{align}

\begin{figure}
    \centering
    \hspace*{-.5cm}\includegraphics[scale=.101]{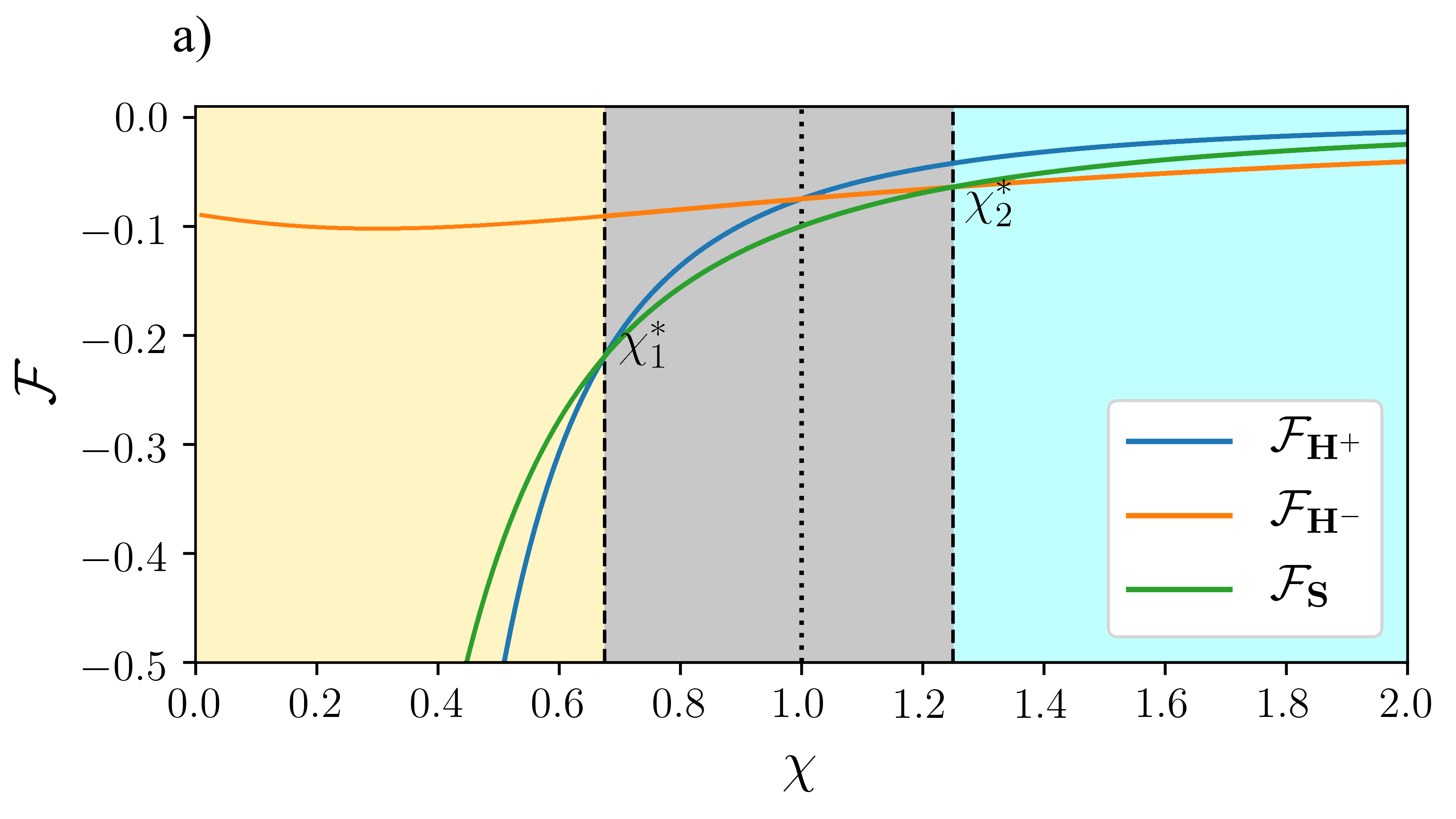}\\
    \hspace*{-.1cm}\includegraphics[scale=.101]{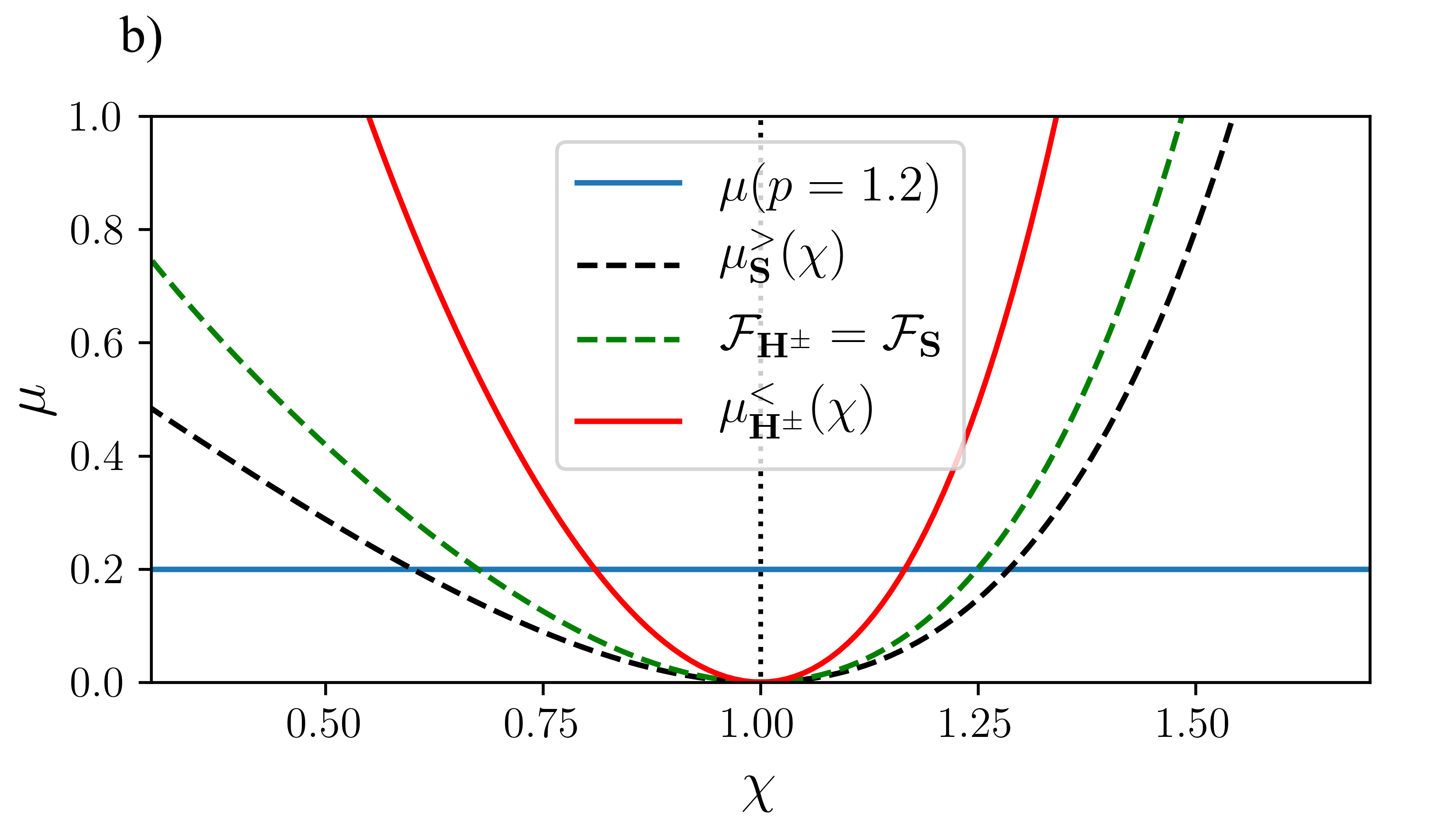}
    \caption{(a) Lyapunov functionals for the three phases at $p=1.2$: $\mathcal{F}_{\mathbf{H}^+}(\chi)$ (blue), $\mathcal{F}_{\mathbf{H}^-}(\chi)$ (orange), $\mathcal{F}_{\mathbf{S}}(\chi)$ (green). The resulting minimum determines the observed self-structured phase while $\chi^*_1$ and $\chi^*_2$ identify the boundaries in Fig.~\ref{pd20}. Note that $\mathcal{F}_{\mathbf{H}^+} = \mathcal{F}_{\mathbf{H}^-}$ for $\chi = 1$ (dotted line).  (b) Critical $\mu^>_{\mathbf{S}}$ and $\mu^>_{\mathbf{H}^\pm}$ (dashed-black/red lines) and phase boundaries (dashed-green) as functions of $\chi$. Intersections with $\mu$ (blue) determine the  size of the $\mathbf{S}/\mathbf{H}$ competition region.}
    \label{plots}
\end{figure}
The Lyapunov functional assumes the following quartic form, as in the weak crystallization scenario \cite{brazovskii1987theory, kats1993weak}:
\begin{equation}
\begin{split}
    \mathcal{F}[\{A_i\}] =& -\mu \sum^3_{i=1} |A_{i}|^2 -\lambda\left( A^*_{1}A^*_{2}A^*_{3} +\textrm{c.c.}\right)+\\
    &\frac{\gamma_2}{2}\sum^3_{i,j=1} |A_{i}|^2|A_{j}|^2 + \frac{\gamma_1}{2}\sum^3_{i=1} |A_{i}|^4,
\label{lyap}
\end{split}
\end{equation}
where $i=1,2,3$ and $i\neq j$.  Non-zero $\lambda$ implies that self-structuring is a first-order phase transition. We obtain the Lyapunov functional for the three phases $\mathcal{F}_{\mathbf{H}^\pm}$ and $\mathcal{F}_{\mathbf{S}}$, and compute the corresponding minimum as a function of $\chi$, shown in Fig.~\ref{plots}(a) \cite{Note1}. In addition we have the critical points \cite{Ciliberto1990}:
\begin{equation}
    \mu^>_{\mathbf{S}} = \frac{\lambda^2
 \gamma_2}{(\gamma_1-\gamma_2)^2}, \quad \mu^<_{\mathbf{H}^\pm} = \frac{\lambda^2(2\gamma_2 + \gamma_1)}{(\gamma_1-\gamma_2)^2},
\label{crit}
\end{equation}
representing the lower $\mathbf{S}$ and the higher $\mathbf{H}^\pm$ stability limits respectively. We overall single out six values of $\chi$, as shown in Fig.~\ref{pd20}. A first couple $\chi^*_{1,2}$  arises from the intersections $\mathcal{F}_{\mathbf{H}^+}(\chi^*_1) = \mathcal{F}_{\mathbf{S}}(\chi^*_1)$ and $\mathcal{F}_{\mathbf{H}^-}(\chi^*_2) = \mathcal{F}_{\mathbf{S}}(\chi^*_2)$ (Fig.~\ref{plots}(a)) and provide phase boundaries in good agreement with the observed ones in Fig.~\ref{pd20} (dashed-green lines). The extremal points in Eq.~(\ref{crit}), shown in Fig.~\ref{plots}(b), yield two other pairs of intersections $\chi^{\mathbf{S}}_{1,2}$ and $\chi^{\mathbf{H}}_{1,2}$, delimiting the $\mathbf{S}/\mathbf{H}$ competition regions around $\chi = 1$ (dashed-black/solid-red lines).
Furthermore, at the same point, the system (displaying $\mathbf{S}$ states) recovers \textit{inversion symmetry} (IS) whereas the $\mathbf{H}^+$ and $\mathbf{H}^-$ states break IS (but are inversion-symmetric to each other). This phenomenon is known for dissipative pattern formation \cite{busse78,malomed87,Ciliberto1990}. The highly interesting feature here is that such a recovery results from a self-tuning depending on the interaction strength $\chi$, while,  otherwise, it typically results from different boundary conditions (e.g.\ in Maragoni compared to Rayleigh-Bénard convection \cite{busse78}), symmetry-breaking external fields \cite{kresic19} or polarization imbalances \cite{Scroggie1996,aumann97}, and strong changes in the homogeneous solution \cite{tlidi93,tlidi94,firth94c,neubecker95}.
\begin{figure}
    \centering
    \hspace{-.35cm}\includegraphics[scale = .075]{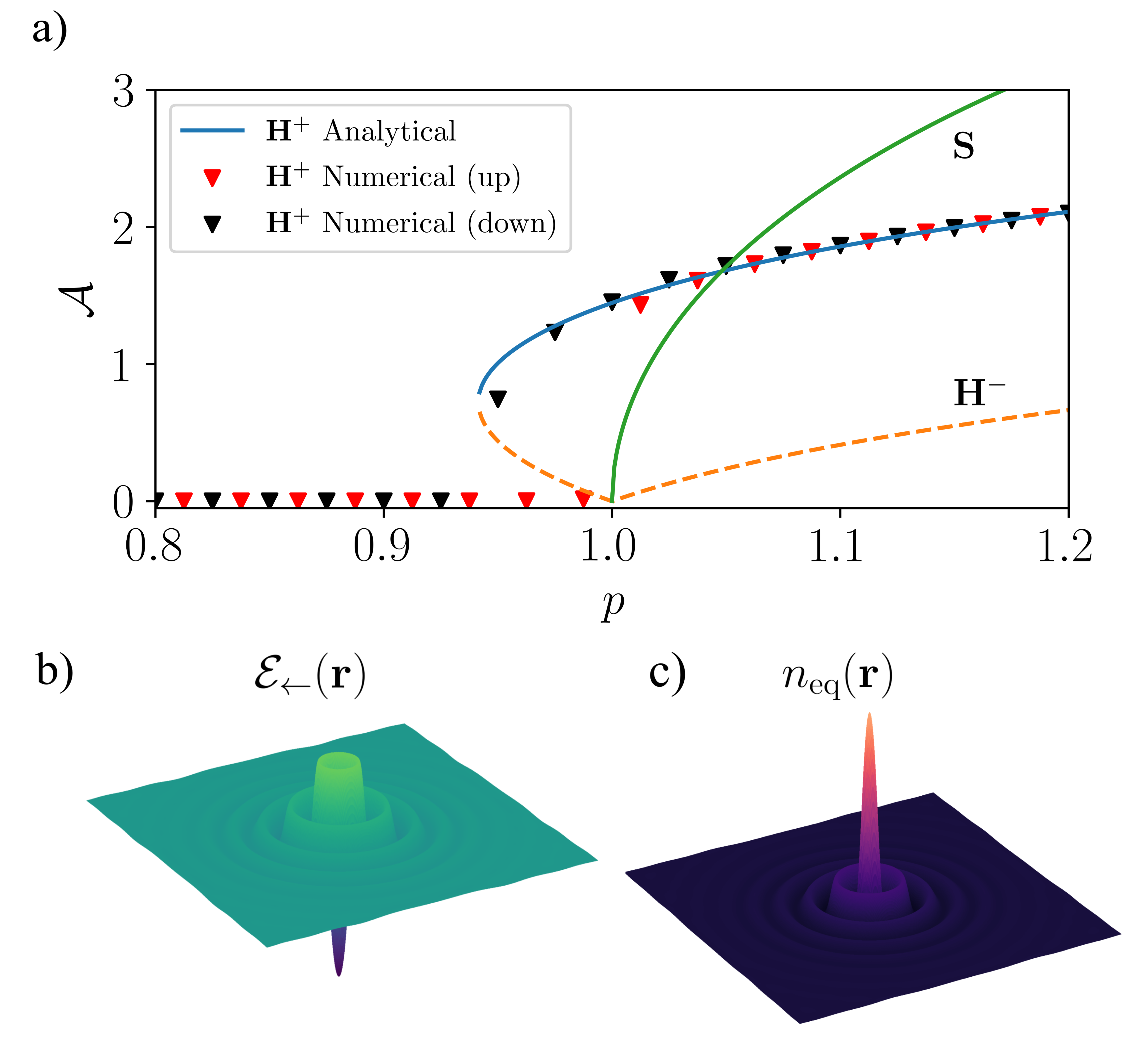}
    \caption{Optomechanical solitons for blue detuning. (a) Amplitude of the $\mathbf{H}^\pm$, $\mathbf{S}$ branches as functions of $p$ for $\chi \approx 0.31$ ($b_0 = 50, \Delta = 80, \sigma\approx 78.3$), plotted together with the numerical amplitude (black/red triangles). (b-c) Dark backwards intensity and bright density profiles at $p=0.98$.}
    \label{soliton1}
\end{figure}

A second intriguing consequence of the optomechanical nonlinearity, elucidated by the AEs, is the possibility of exciting light-density spatial solitons when $\lambda \neq 0$ \cite{ackemann2009fundamentals, Tesio2013}. Indeed, as a universal feature of the AEs (\ref{GLEs}), the $\mathbf{H}^\pm$ branches display subcriticality, i.e. they are stable in a negative range $\mu_{\textrm{SN}}<\mu <0$, originating in a saddle-node bifurcation at:
\begin{equation}
    \mu_{\textrm{SN}} = -\frac{\lambda^2}{4(\gamma_2+2\gamma_1)}.
\end{equation}
\begin{figure}
    \centering
    \hspace{-.4cm}\includegraphics[scale=.105]{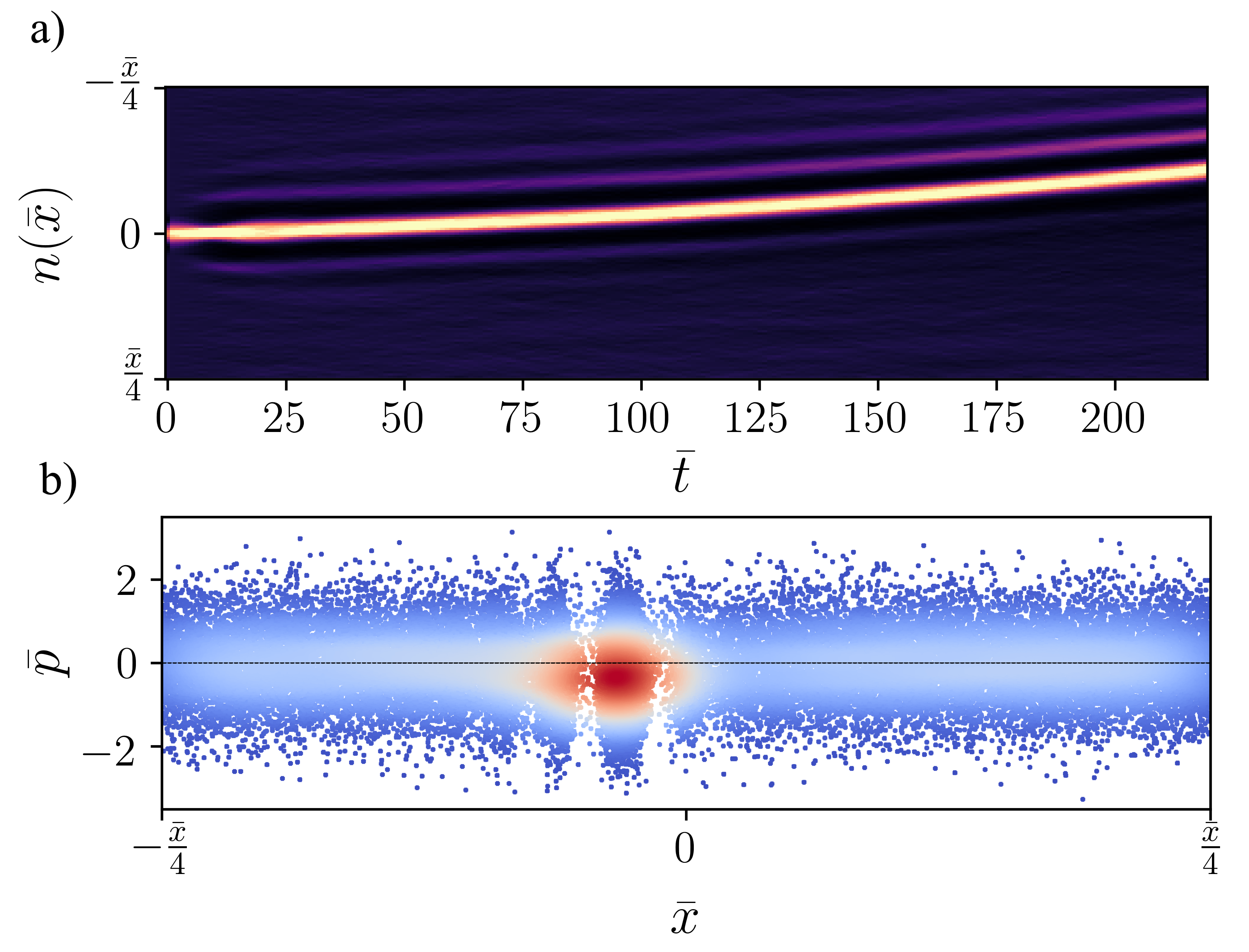}
    \caption{1D angular dynamics simulation of $4 \times 10^4$ atoms with a driving beam possessing OAM (index $l=1$) \cite{Note1}. (a) Density evolution $n(\bar{x},\bar{t})$, numerically reconstructed from particle trajectories. (b) Phase space distribution at $\bar{t} = 120$. }
    \label{particles}
\end{figure}
This is shown for $\Delta >0$ in Fig.~\ref{soliton1}(a), where the stable $\mathbf{H}^+$ branch $A_1=A_2=A_3=\mathcal{A}$ (blue line), computed analytically from the AEs coefficients above, is in good agreement with the numerical amplitude $(\max(n_{\textrm{eq}})-\min(n_{\textrm{eq}}))/2$ for $p \in [0.8,1.2]$ \cite{Note1}. The stability of $\mathbf{H}^+$ for $\chi < 1$ allows for the existence of a spatial feedback soliton characterized by a dark intensity profile  $|\mathcal{E}_{-}(\rp)|^2$, which serves as a self-sustained trap for a bright density peak, as displayed in Fig.~\ref{soliton1}(b)-(c).
Controlling soliton motion via external phase gradients enables atomic transport applications \cite{Firth1996a, yao2019control}. We address that by means of 1D particle dynamics simulations, where parameters are tuned in order to match those in Fig.~\ref{soliton1} in the thermodynamic limit \cite{Note1}. Assuming periodic boundary conditions, the atoms are effectively confined in an annular trap and, thus, a linear phase on the input field corresponds to the 1D angular equivalent of orbital angular momentum (OAM) \cite{Baio2020}. The density profile is shown in Fig.~\ref{particles}(a) where, after a transient behaviour, the atoms initially prepared in a density peak reach steady state angular drift, induced by OAM. This is illustrated by the phase space distribution in Fig.~\ref{particles}(b), where the non-zero momentum of the trapped region is visible.

In summary, we have demonstrated transverse optomechanical self-structuring to hexagonal, stripe and honeycomb phases in cold atomic clouds subject to optical feedback. Focusing on a simple model of overdamped motion, we pointed out that the Kerr picture of the optomechanical nonlinearity fails to capture structural transitions among hexagons, stripes, and honeycombs, depending on the coupling strength. Indeed, in that case, only the second addend in Eq.~(\ref{coef12}) arises as the resulting nonlinearity involves the intensity alone, resulting in a pure quadratic dependence on the susceptibility  \cite{DAlessandro1992}. By contrast, the optomechanical nonlinearity involves the transport generating product $n(\mathbf{r},t)\nabla_\perp s(\mathbf{r},t)$ \cite{Tesio2012}, so that the mixing of linear terms from both factors gives rise to a shifted quadratic dependence on $\chi$, becoming effective-Kerr for $\chi > 1$ only. For $\chi = 1$, the system is inversion symmetric,  undergoing a structural transition to a stripe state.

Structural phase transitions received recent attention in the context of dipolar supersolids  \cite{Zhang2019}, and driven Bose-Einstein condensates \cite{zhang2020pattern}. To our knowledge, there is no experimental confirmation of a stripe-like inversion symmetric supersolid phase, as current experiments address only quasi-1D configurations \cite{guo2019low, natale2019excitation}. In light of the similarity between the free energy discussed here and the energy functional for a dipolar condensate, we conjecture the existence of an intermediate, inversion-symmetric, supersolid phase (stripe or square-like). The proposed scheme provides relative ease of experimental implementation of 2D symmetry-breaking phenomena in cold atoms \cite{Labeyrie2014a}, and quantum degenerate gases \cite{Robb2015, Zhang2019}. The overdamped limit under scrutiny here simplifies the analytical treatment but the present phase selection occurs in the Hamiltonian case, i.e., without optical molasses \cite{Note1}. Finally, the existence of optomechanical feedback solitons motivates us to explore analogues in quantum degenerate gases \cite{Robb2015}, in connection with the concept of quantum droplets \cite{PhysRevLett.116.215301, zhang2020self}.

\begin{acknowledgments}
All authors acknowledge financial
support from the European Training Network ColOpt, which
is funded by the European Union (EU) Horizon 2020 program
under the Marie Skłodowska-Curie Action, Grant Agreement
No. 721465. We are grateful to G. Labeyrie and R. Kaiser for numerous discussions and the fruitful collaboration on optomechanical self-organization.
\end{acknowledgments}

\bibliography{short, paperSFM,thorstenlit}
\bibliographystyle{apsrev}
\end{document}